\documentclass[12pt]{emulateapj}
\usepackage{xspace,graphics,graphicx}
\newcommand\msun{\ensuremath{M_\sun}\xspace}
\newcommand\teff{\ensuremath{T_{\rm eff}}\xspace}
\newcommand\logg{\ensuremath{\log g}\xspace}

\newcommand\mch{\ensuremath{M_{\mathrm Ch}}\xspace}

\slugcomment{Accepted for Publication in the Astronomical Journal} 
\shorttitle{Candidate double degenerates in NGC 6633}
\shortauthors{Williams et al.}

\begin{document}
\title{Time-series spectroscopy of two candidate double degenerates in the open cluster NGC 6633 \footnote{The data presented herein were obtained at the W.M. Keck Observatory, which is operated as a scientific partnership among the California Institute of Technology, the University of California and the National Aeronautics and Space Administration. The Observatory was made possible by the generous financial support of the W.M. Keck Foundation. }}

\author{Kurtis A.~ Williams}
\email{Kurtis.Williams@tamuc.edu}
\affil{Department of Physics \& Astrophysics, Texas A\&M University-Commerce, P.O. Box 3011, Commerce, TX, 75429, USA}
\author{Donald Serna-Grey\altaffilmark{1}}
\altaffiltext{1}{REU Participant, Department of Physics \& Astrophysics, Texas A\&M University-Commerce, Commerce, TX}
\affil{Department of Astronomy, University of Washington, Box 351580, Seattle, WA, 98195, USA}
\author{Subho Chakraborty\altaffilmark{2}}
\affil{Department of Physics \& Astrophysics, Texas A\&M University-Commerce, P.O. Box 3011, Commerce, TX, 75429, USA}
\altaffiltext{2}{Current address: The Graduate Center, City University of New York, New York, NY, USA}
\author{A.~Gianninas}
\author{Paul A.~Canton}
\affil{Homer L. Dodge Department of Physics and Astronomy, University of Oklahoma, 440 W.~Brooks St., Norman, OK, 73019, USA}

\begin{abstract}
Type Ia supernovae are heavily used tools in precision cosmology, yet we still are not certain what the progenitor systems are.   General plausibility arguments suggest there is potential for identifying double degenerate Type Ia supernova progenitors in intermediate-age open star clusters.  We present time-resolved high-resolution spectroscopy of two white dwarfs in the field of the open cluster NGC 6633 that had previously been identified as candidate double degenerates in the cluster. However, three hours of continuous observations of each candidate failed to detect any significant radial velocity variations at the $\gtrsim 10$ km s$^{-1}$  level, making it highly unlikely that either white dwarf is a double degenerate that will merge within a Hubble Time.  The white dwarf LAWDS NGC 6633 4 has a  radial velocity inconsistent with cluster membership at the $2.5\sigma$ level, while the radial velocity of LAWDS NGC 6633 7 is consistent with cluster membership.  We conservatively conclude that LAWDS 7 is a viable massive double degenerate candidate, though unlikely to be a Type Ia progenitor.  Astrometric data from GAIA will likely be needed to determine if either white dwarf is truly a cluster member.
\end{abstract}
\keywords{binaries: close --- binaries: spectroscopic  --- open clusters and associations: individual (NGC 6633) --- supernovae: general --- white dwarfs}

\section{Introduction}

The double-degenerate (DD) channel for Type Ia supernovae proposes that two carbon-oxygen 
core white dwarfs (WDs) with a total mass at or above the Chandrasekhar mass, \mch, will spiral inwards
due to gravitational radiation, merge, and explode \citep{1984ApJS...54..335I}.    The degree to 
which the DD channel contributes to Type Ia supernovae is still a matter of vociferous debate \citep[e.g.,][]{2014ARA&A..52..107M,2015Natur.519...63S,2015arXiv150301739G}.  We present observations of two candidate DDs in an intermediate-age open star cluster that, while not constraining the DD channel any further, illustrate the potential utility of WDs in open star clusters as probes of the DD channel. 
%to help resolve this debate and to investigate two promising DD candidates in the field of the 
%open cluster \object{NGC 6633}.

Observational and theoretical arguments from several groups suggest that some fraction of Type Ia 
supernovae come from a ``prompt'' population, with supernova onset $\lesssim 500$ Myr after a 
star formation event, as well as a second progenitor population with a much longer delay before 
detonation \citep[e.g.,][]{Barris2006,Mannucci2005,Scannapieco2005a,Sullivan2006,Maoz2010b,Maoz2011a}. Therefore, 
stellar populations with ages $\lesssim 500$Myr should contain more prompt Type Ia progenitors, 
whatever these progenitor systems may actually be.

Regardless of the delay time, the DD Type Ia progenitor channel requires at least one of the WDs to 
have a mass $\gtrsim M_{\rm Ch} / 2\approx 0.7\msun$.  Based on the semi-empirical initial-final 
mass relation, such a WD will result from a progenitor star with $M_{\rm init}\gtrsim 2.8 \msun$ 
\citep[e.g.,][]{2008MNRAS.387.1693C,2009ApJ...693..355W,Dobbie2009,Dobbie2012,2012ApJ...746..144Z,2015ApJ...807...90C} and with a nuclear lifetime of $\lesssim 
560$ Myr, assuming metallicity $Z=0.019$ \citep{Marigo2008}, consistent with the delay time discussed in the previous paragraph.  Given the known scatter in the initial-final mass relation around $\sim 3\msun$, we emphasize that these numbers are intended to be estimates only. 

The upper limits on orbital periods producing a DD merger within a Hubble time require that Type 
Ia progenitors must go through a common envelope phase.  It has been observed that the binary 
frequency of stars rises from late-type B stars through O stars ($M_*\gtrsim 3\msun$), and that these 
massive binaries are far more likely to have mass ratios closer to unity than expected from random 
chance \citep[e.g.,][]{Lucy1979,Pinsonneault2006,Duchene2013}.  Since at least one of the 
members of a DD must have originated from a star of spectral type late B or earlier in order to 
produce a WD of mass $\geq 0.7\msun$, it is plausible that both degenerates in any DD Type Ia 
progenitor system evolved from stars with main sequence masses $\gtrsim 3\msun$, which have 
nuclear lifetimes of $\lesssim 500$ Myr.

Taken together, these models and observations suggest that DD Type Ia supernova progenitors, if 
they exist, could form within $\sim 500$ Myr of the onset of star formation. Therefore, a simple stellar 
population of age $\lesssim 500$ Myr, such as open star clusters with main sequence turnoff 
masses $\gtrsim 3\msun$, may be relatively fertile ground for hunting DD Type Ia progenitor 
systems.  If the DD systems have not already merged and detonated as prompt Type Ia supernovae 
prior to our observations, the individual WD components should be relatively bright and hot, and 
thus straightforward to identify in intermediate-age open star clusters with existing ground-based 
facilities.

\subsection{Two Candidate Double Degenerates in the Field of NGC 6633}

For the past several years, we have engaged in studies of WDs in intermediate-age open clusters, 
with goals of observationally constraining the relationship between WD masses and their 
progenitor star masses (the initial-final mass relation) as well as the maximum mass of 
WD progenitors \citep{Williams2002,Williams2004,Liebert2005,Williams2007,2008AJ....135.2163R,
2009ApJ...693..355W,Dobbie2009,Dobbie2012,Liebert2013}.

One necessary step in this process is identifying confirmed WDs as cluster 
members.  The vast majority of the WDs we have discovered do not have existing proper motion 
measurements precise enough for cluster membership determinations.  Instead, we have 
constrained cluster membership via a version of spectroscopic parallax described in detail in \citet{Williams2007} and \citet{2009ApJ...693..355W}.   In summary, we obtain absolute magnitudes for each WD via \teff and \logg spectral fits of observed spectra to model atmospheres combined with WD evolutionary models that include the mass-radius relation.  In this paper, we use the color and model calculations from \citet{2006AJ....132.1221H,2006ApJ...651L.137K,2011ApJ...730..128T}; and \citet{2011ApJ...737...28B}\footnote{Available online at\\ \url{http://www.astro.umontreal.ca/~bergeron/CoolingModels/}}.  

We then subtract the apparent 
magnitude obtained via optical photometry from the derived absolute magnitude to obtain the apparent distance modulus of the WD. If the apparent distance modulii of the WD and star cluster  are consistent, and if the 
WD cooling age is less than the cluster age, then we identify the WD as a likely cluster member.  
This greatly reduces field contamination in the WD sample, but it precludes the inclusion of WD-WD 
binary systems if the secondary star contributes any significant flux.  

In \citet{Williams2007}, we confirmed the presence of 13 WDs in the field of the open cluster \object{NGC 
6633}.  NGC 6633 has an age of $\approx 560$ Myr \citep{Williams2007}, a slightly subsolar metallicity $[{\rm Fe}/{\rm H}]\approx -0.1$, a distance modulus $(m-M)_0 = 8.01\pm 0.09$, and moderate reddening $E(\bv)=0.165\pm 0.011$ \citep{2002MNRAS.336.1109J}.  

Based on the \citet{Williams2007} analysis, only one WD  (\object{LAWDS NGC 6633 27}) is consistent with the apparent distance modulus of $(m-M)_V=8.52\pm 0.10$ (assuming $R_V=3.1$).  However, two 
additional WDs (\object{LAWDS NGC 6633 4} and \object{LAWDS NGC 6633 7}, hereafter LAWDS 
4 and LAWDS 7, respectively) are significantly more massive ($\geq 0.8\msun$) than the mode of 
the field WD mass distribution \citep[$\approx0.65\msun$][]{Tremblay2009}.  Additionally, both 
have fluxes almost exactly a factor of two higher than would be expected for cluster members, i.e., apparent distance moduli 0.75 mag less  than the cluster.  Astrometric and photometric data for these two objects are given in Table \ref{tab.pos}.

\begin{deluxetable*}{lllcccc}
\tablecolumns{7}
\tablewidth{0pt}
\tablecaption{Astrometric and Photometric Parameters of the Candidate Double Degenerates\label{tab.pos}}
\tablehead{\colhead{Object} & \colhead{R.A.} & \colhead{Decl.} & \colhead{$\mu_{\rm RA}$} & \colhead{$\mu_{\rm Decl}$} & \colhead{$V$} & \colhead{$\bv$} \\
 & (J2000) & (J2000) & (mas/yr) & (mas/yr) & (mag) & (mag)}
\startdata
LAWDS 4 & 18 27 10.4 & 06 26 15.7 & $-8.0\pm 2.0$ & $6.0\pm 5.0$ & $18.82\pm 0.02$ & $0.16\pm 0.03$ \\
LAWDS 7 & 18 27 49.9 & 06 20 51.8 & \nodata & \nodata & $19.27\pm 0.02$ & $0.23\pm 0.03$ \\
\enddata
\tablecomments{Proper motions are from \citet{2003AJ....125..984M}; other data are from \citet{Williams2007}. }
\end{deluxetable*}

Proper motion cluster memberships of these 
two stars using astrometric data from the USNO-B catalog \citep{2003AJ....125..984M} are 
inconclusive. The proper motion vector $(\mu_{\rm RA}, \mu_{\rm Decl})$ of NGC 6633 is $(0.1\pm 0.22, -2.00\pm 0.24)$ mas yr$^{-1}$ \citep{2005A&A...438.1163K}, while the surounding fields stars have a mean proper motion vector of $(-2.06, -5.58)$ mas yr$^{-1}$ with a dispersion of $(20.5, 20.8$) mas yr$^{-1}$ \citep{2014A&A...564A..79D}.
LAWDS 4 has a right ascension proper motion formally inconsistent with the cluster at the $ 4\sigma$ level, but USNO-B catalog proper motions of bright QSOs show an rms dispersion of 6.18 mas yr$^{-1}$ in right ascension near the celestial equator \citep{2004AJ....127.3034M}, essentially negating this result.  LAWDS 7 has no detected proper motion in the USNO-B catalog.  These results are not surprising as proper motion is found to bee a poor discriminator for faint stars in this field; the mean cluster tangential velocity is only $\sim 1.5$ km s$^{-1}$ compared to the field  \citep{1997MNRAS.292..177J}.

Simulations of star cluster evolution suggest binary WD cooling sequences should exist at roughly 
this overluminosity \citep[e.g.,][]{Hurley2003,Geller2013}, and features in the WD luminosity function of the rich open 
cluster NGC 6791 can be interpreted as a binary cooling sequence 
\citep[e.g.,][]{GarciaBerro2011}, though a He-core WD cooling sequence also explains these 
observations \citep{Kalirai2007}. We therefore suggested in \citet{Williams2007} that LAWDS 4 and 
LAWDS 7 may be double degenerates.  

Since our initial publication, other open cluster WD studies have flagged WDs with distance moduli 
$\leq 0.75$ mag foreground to their cluster as candidate DDs \citep[e.g.][]{2008ApJ...676..594K,
2008AJ....135.2163R}, though none of the proposed candidates are so enticingly close to the 0.75 mag over-luminosity often indicative of binary systems.  If LAWDS 4 and LAWDS 7 are equal-mass DDs, each system's total mass is super-Chandrasekhar.  
Based on these arguments, we proposed for and 
received time at the Keck Observatory through the NOAO TSIP program (NOAO proposal 
08A-0124) to perform follow-up spectroscopy on these two candidate double degenerate systems.

\section{Revised White Dwarf Parameters}\label{sec.params}

Our WD parameters in \citet{Williams2007} are based on Balmer line
fits using slightly modified versions of synthetic, pure-H atmospheres
used in \citet{Finley1997} and graciously provided by D.~Koester at
the time.  Since that time, the inclusion of additional physics and of
new calculations of quantum effects and Stark broadening has led to
small but systematic shifts in \teff and \logg determinations for DA
WDs \citep{Tremblay2009,Koester2009,Koester2010}.  Under the
assumption that the resulting new atmospheric models and parameters
are more accurate than the earlier models, we have recalculated the
surface gravity and effective temperatures for LAWDS 4 and LAWDS 7.

We simultaneously fit the Balmer lines to model spectra using the spectroscopic technique
developed by \citet{Bergeron1992}; we applied the recent innovations and techniques presented in \citet{Gianninas2011} and references therein. We first
normalized each individual Balmer line such that the continuum level was set to unity in
both the observed and model spectra. We compared the normalized observed spectra with the normalized synthetic spectra after convolving them with a Gaussian function approximating the instrumental profile.  Using the steepest descent nonlinear least-squares method of Levenberg-Marquardt \citep{1992nrfa.book.....P}, we determined \teff and \logg from the grid of model spectra.  
We did not apply the 3D convective model corrections of \citet{2013A&A...552A..13T}, as LAWDS 4 and LAWDS 7  are sufficiently hot that their atmospheres are radiative, and the corrections would be negligible.  

The new WD parameters, along with the derived WD masses,  cooling times, and distance moduli
are given in Table~\ref{tab.newfits}.  A color-magnitude diagram for the cluster, with the two DD candidates indicated by open squares, can be found in Figure 6 of \citet{Williams2007}.  Figures 8a and 8b of \citet{Williams2007} show how the physical parameters of these two WDs compare to those expected for cluster members with the same apparent magnitudes; the new physical parameters shift these stars to slightly higher temperatures.  The newer WD masses 
and distance moduli are consistent with the earlier published values, and these two WDs 
still stand out as candidate DDs.

\begin{deluxetable*}{lccccccc}
\tablecolumns{8}
\tabletypesize{\footnotesize}
\tablewidth{0pt}
\tablecaption{Revised Atmospheric Parameters for LAWDS 4 and LAWDS 7\label{tab.newfits}}
\tablehead{\colhead{Object} & \colhead{\teff} & \colhead{\logg} & \colhead{$M_{\rm WD}$} & \colhead{$\tau_{\rm cool}$} & \colhead{$M_V$} &\colhead{$(m-M)_V$} & \colhead{$\Delta[(m-M)_V]$}\tablenotemark{a} \\
 & (K) & & (\msun) & (Myr) & (mag) & (mag) & (mag) }
\startdata
LAWDS 4 & $21800\pm 350$ & $8.29\pm 0.05$ &$0.80\pm 0.03$ & 98.8 & $11.07\pm 0.09$ & $7.74\pm 0.09$ & $-0.78\pm 0.13$ \\
LAWDS 7 & $19360\pm 310$ & $8.42\pm 0.05$ & $0.88\pm 0.03$ & 187 & $11.48\pm 0.09$ & $7.78\pm 0.09$ & $-0.74\pm 0.13$ \\
\enddata
\tablenotetext{a}{Difference of the WD and NGC 6633 distance modulii}
\end{deluxetable*}

\section{High-Resolution Spectroscopy}

WDs of spectral type DA often exhibit sharp, narrow line cores superimposed on the broad Balmer 
absorption lines.  These cores are attributed to non-local thermodynamical equilibrium (NLTE) 
conditions in the WD atmosphere \citep{1973A&A....25...29G}. The goal of our new observations 
was to obtain time-series spectroscopy with sufficient resolution to detect the NLTE core of H$\alpha
$, similar to methods employed by the European Southern Observatory SN Ia Progenitor Survey 
\citep{Napiwotzki2001}.  Since we had only one short summer night for observations, 
we observed each target for only 3 h.  Although DD Type Ia progenitors can have orbital periods of 
up to 12 hours, detection of even a partial orbit should be sufficient to warrant additional observing 
time.  

We obtained data with the Echellette Spectrograph and Imager \citep[ESI,][]{Sheinis2002} on Keck II 
on UT 2008 Jul 3.  We operated the instrument in echelle mode with a slit width of 0\farcs75.  
We measured the velocity resolution using the full-width at half-maximum (FWHM) of the 5577\AA\  
night sky line at 55.7 km/s, consistent with the stated instrumental resolution of 55.9 km/s ($\approx 
5$ pix).  The weather was photometric with steady seeing of 0\farcs6.  For exposures of LAWDS 7, 
one mirror segment did not align properly, leading to a faint second image in the spatial direction; 
we ignored this second image in our further analysis.

We varied exposure times for LAWDS 4 in hopes of detecting any possible velocity smearing 
resulting from orbital periods $\lesssim 20$ minutes.  This resulted in signal-to-noise (S/N) ratios 
per pixel ranging from 13 (600 s exposures) to 21 (1200 s exposures) in the continuum surrounding 
the the H$\alpha$ line.  Exposure times for LAWDS 7 were held constant after the first exposure due 
to the lower flux from this fainter star, with ${\rm S/N}\approx 18 \, \mathrm{pix}^{-1}$.    Details on 
each target exposure are presented in Table~\ref{tab.wb4} for LAWDS 4 and Table~\ref{tab.wb7} for 
LAWDS 7.

We reduced the data using the ESIRedux package of \citet{Prochaska} and followed the basic 
recipe for reductions given in the ESIRedux Cookbook\footnote{Available at\\ \url{http://www2.keck.hawaii.edu/inst/esi/ESIRedux/esi_cookbook.html}}. ESIRedux applies corrections to the 
data from bias images and flat fields, extracts the object spectrum, subtracts background sky, 
wavelength calibrates the extracted spectra, and uses night sky lines to correct the wavelength zero 
points for any instrumental flexure remaining in each individual exposure.  

We applied bias frames and HgNe and CuAr arc lamp solutions obtained during the run.  ESIRedux 
was not able to use our internal flat fields, so we used an archived dome flat provided with the 
ESIRedux code for flat fielding; no obvious flat fielding errors remained in the spectra.  We also 
used an archived map of the order curvature provided with the ESIRedux code for the spectral 
tracing and extraction; again, the results are satisfactory.

We used exposures of \object{Feige 67} to apply a relative 
flux correction.  The orders were stitched together to produce a one-dimensional output spectrum; 
wavelengths are calibrated to vacuum wavelengths. The resulting spectra are of high quality for 
$4000 {\rm \AA} \leq \lambda \leq 8000 {\rm \AA}$, though residual noise from the regions 
of order overlap were visible.  We confirmed the wavelength solutions by measuring centroids of strong night-sky emission lines; these centroids are consistent with zero velocity with a dispersion of less than 1 
km s$^{-1}$.  Due to this, we did not apply corrections from a radial velocity standard star.  Further, as we were interested primarily in the Balmer absorption lines, we did not apply telluric corrections to the data.

\begin{figure}
\includegraphics[angle=270,width=\columnwidth]{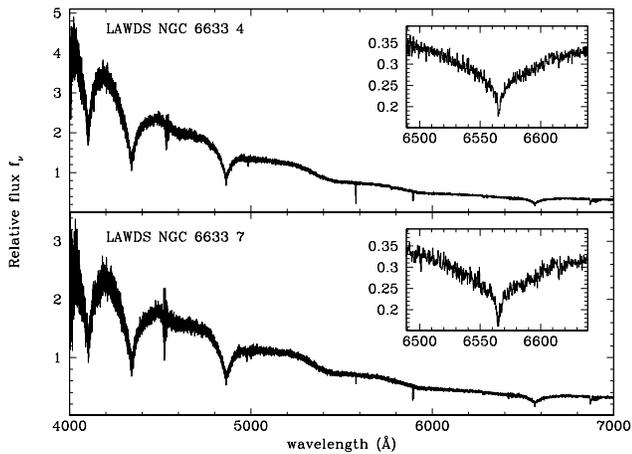}
\caption{Coadded ESI spectra for LAWDS 4 (top) and LAWDS 7 (bottom).  Insets show the central 
regions of the H$\alpha$ absorption lines and the NLTE cores.  Narrow interstellar Na D lines and 
telluric B-band absorption are visible, as are residuals from 5577\AA\  night sky line subtraction.  
Most other high-frequency features (such as the wiggles at 4500\AA\  and 6400\AA) are due to the order overlap regions.  Both spectra qualitatively 
look like typical single DA white dwarfs.\label{fig.combspectra}}
\end{figure}

Figure~\ref{fig.combspectra} shows the co-added spectra from all
exposures of each target.   In addition to the broad Balmer absorption lines, narrow Na D lines are observed; these are presumably interstellar in origin.

\begin{figure}
\begin{center}
\includegraphics[width=0.9\columnwidth]{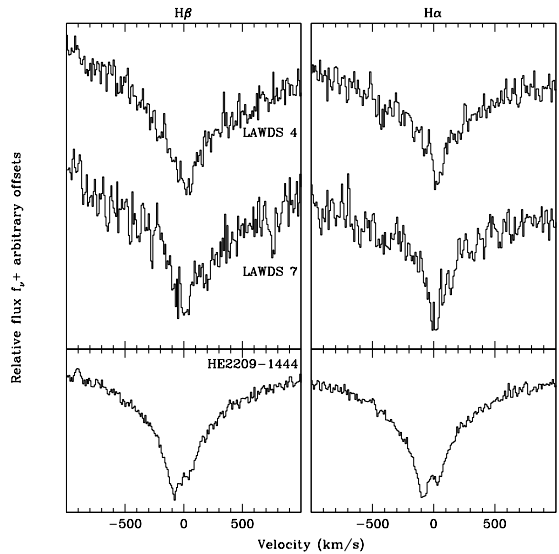}
\end{center}
\caption{The H$\beta$ (left) and H$\alpha$ (right) absorption line centers for LAWDs 4 and 7 (top 
panel) and the known double degenerate HE 2209$-$1444, obtained with the same instrumental 
configuration at a random phase angle.  The two NLTE cores are visible in HE 2209$-$1444, but 
only a single line core is visible in LAWDS 4.  LAWDS 7 has a feature suspiciously similar to the second cores in HE 2209$-$1444, but we cannot consider this significant; see the discussion in the text. 
\label{fig.he2209}}
\end{figure}

Figure~\ref{fig.he2209} shows the central regions of the H$\alpha$ and H$\beta$ lines for LAWDS 
4 and LAWDS 7, as well as of the known short period DD \object{HE 2209$-$1444}, the data for 
which were obtained on the same night with the same instrumental configuration, but at a random 
phase angle.  This illustrates what a double-lined spectroscopic binary would look like in our data, 
given sufficient signal-to-noise.  Visual inspection showed evidence for only one NLTE core in the H$\alpha$ line of LAWDS 4.  

The H$\alpha$ and H$\beta$ cores of LAWDS 7 exhibit one obvious 
NLTE core and evidence of a second absorption feature redward of the 
strong core.  This second feature is qualitatively similar to the weaker NLTE core in the spectrum of  HE 2209$-$1444.  However, the centroids of the redder features differ in the two lines -- in H$\alpha$, the second dip is at a velocity of $\approx +130$ km s$^{-1}$, while in H$\beta$ it appears at a velocity of $\approx 200$ km s$^{-1}$.  Further, features with similar depths appear elsewhere in the lines (e.g., at +250 km s$^{-1}$ in LAWDS 7 H$\beta$, and at $-80$ km s$^{-1}$ in LAWDS 4 H$\alpha$).   Therefore we consider these features intruiging, but they are neither significant nor convincingly a second NLTE core.  Further, given the 3-hour total integration time of the coadded spectra, one would expect radial velocity smearing in the line cores, but such smearing is not consistent with the measured FWHMs.

We note that the NLTE cores can vary in strength or even appear in emission depending on the 
atmospheric parameters and structures.  The lack of a 2nd NLTE core does not, therefore, preclude 
the system from being a double-lined spectroscopic binary if such emission has filled in the line core.  

\subsection{Time-series radial velocities}\label{sec.hires}
We search for radial velocity variations using the individual spectral exposures listed in Tables 
\ref{tab.wb4} and \ref{tab.wb7}.  Unfortunately, the NLTE H$\alpha$ core is weak in individual 
exposures.  We should have expected this given the moderate \teff of our candidates; the H$\alpha
$ NLTE core becomes less well defined at higher \teff in DA WDs.  

\begin{deluxetable}{lccc}
\tablecolumns{4}
\tablecaption{Observations and Radial Velocity Data for LAWDS 4\label{tab.wb4}}
\tablewidth{0pt}
\tablehead{\colhead{Exposure} & \colhead{$T_{\rm mid}$} & \colhead{$t_{\rm exp}$}  & \colhead{$v_{\rm rel}$} \\
Number & (HJD$-2454650.5$) & (s) & (km s$^{-1}$)  }
\startdata
  1 & 0.29067 &   900 & $ -17.9\pm 7.6$ \\
  2 & 0.30173 &   900 & $ -24.4\pm 7.2$ \\
  3 & 0.31455 &   900 & $ -24.3\pm 7.7$ \\
  4 & 0.32915 &   900 & $ -24.4\pm 8.4$ \\
  5 & 0.34372 & 1200 & $ -22.3\pm 7.7$ \\
  6 & 0.35514 & 1200 & $ -32.3\pm 7.5$ \\
  7 & 0.36276 & 1200 & $ -33.3\pm 7.6$ \\
  8 & 0.37037 &   600 & $ -28.8\pm 6.4$\\
  9 & 0.37799 &   600 & $ -30.2\pm 7.3$ \\
10 & 0.38561 &   600 & $ -13.9\pm 7.9$ \\
11 & 0.39499 &   600 & $ -25.7\pm 7.8$ \\
12 & 0.40617 &   600 & $ -41.5\pm 7.6$ \\ 
\enddata
\tablecomments{Velocities are not absolute, but relative to a model atmosphere and include heliocentric corrections; see Section \ref{sec.hires}}
\end{deluxetable}

\begin{deluxetable}{lccc}
\tablecolumns{4}
\tablecaption{Observations and Radial Velocity Data for LAWDS 7\label{tab.wb7}}
\tablewidth{0pt}
\tablehead{\colhead{Exposure} & \colhead{$T_{\rm mid}$} & \colhead{$t_{\rm exp}$}  & \colhead{$v_{\rm rel}$} \\
Number & (HJD$-2454650.5$) & (s) & (km s$^{-1}$)  }
\startdata
 1 & 0.43018 &          900 & $   -31.5\pm 7.5$ \\
 2 & 0.44304 &       1200  & $   -18.8\pm 7.0$ \\
 3 & 0.45758 &       1200  & $   -33.2\pm 5.1$ \\
 4 & 0.47215 &       1200  & $   -39.9\pm 7.6$ \\
 5 & 0.48668 &       1200  & $   -31.1\pm 8.0$ \\
 6 & 0.50141 &       1200  & $   -24.3\pm 5.4$ \\
 7 & 0.51598 &       1200  & $   -37.2\pm 7.3$ \\
 8 & 0.53051 &       1200  & $   -26.4\pm 3.6$ \\
\enddata
\tablecomments{Velocities are not absolute, but relative to a model atmosphere and include heliocentric corrections; see Section \ref{sec.hires}}
\end{deluxetable}

After we unsuccessfully attempted several methods of  measuring robust radial velocities from the 
H$\alpha$ and H$\beta$ NLTE line cores alone \citep{2013PhDT.......281C}, we changed our approach to utilize larger portions 
of the spectrum.  We used the \emph{fxcor} package in \emph{IRAF}\footnote{IRAF is distributed by 
the National Optical Astronomy Observatory, which is operated by the Association of Universities 
for Research in Astronomy (AURA) under a cooperative agreement with the National Science 
Foundation.} \citep{1993ASPC...52..472F}.  We ran cross-correlations using a variety of continuum 
fitting techniques, Fourier filtering parameters, and wavelength ranges, including the central 
regions of the Balmer absorption lines, the entirety of the lines, and the entire spectrum.  The final 
constraints described in the next paragraph resulted in radial velocity measurements that were the 
most self-consistent and stable against small changes in the task parameters.

We fit the observed and template spectra continua (see below) with a fourth-order cubic spline over 
wavelengths of 4180\AA ~to 7500\AA.  We rejected discrepant wavelength  points using sigma-
clipping rejection of $5\sigma$ for points above the fit spline and $2\sigma$ for points below the 
spline; we ran eight rejection iterations.  This resulted in efficient exclusion of the broad Balmer 
absorption lines as well as noise from poor night sky subtraction and regions of spectral order 
overlap.  Visual inspection of the normalized template spectra suggested an excellent continuum fit, 
while the normalized object spectra often had remaining low-frequency wiggles in the continuum.  
These fluctuations were highly resistant to efforts to remove them, and we were loathe to increase 
the order of the continuum fitting spline lest we inadvertently introduced significant artificial 
residuals into the Balmer line profiles. 

To minimize the effects of this low-frequency residual in the continuum and to reduce any impact of 
cosmic ray hits or poor sky subtraction on the cross-correlation, we Fourier filtered the normalized 
spectrum using a ramp filter rising linearly from zero to one between Fourier wavenumbers $k_1$ 
and $k_2$, and falling linearly back to zero between wavenumbers $k_3$ and $k_4$.  We selected 
$k_1 =6$ and $k_2=9$  based on best practice suggestions by \citet{1999MNRAS.305..259W} and 
\citet{2009arXiv0912.4755A}.  The low-frequency object spectrum continuum fitting residuals were 
effectively removed by this filtering.  We set $k_3=3000$ (roughly one spectral resolution element) 
and $k_4 = 6000$ (roughly three pixels) to remove any high-frequency signals that could not be 
present in the intrinsic WD spectrum.

Finally, our cross-correlation focused on the spectral regions within 25\AA~ of the rest-wavelength 
H$\alpha$ and H$\beta$ line centroids.    Significantly larger spectral regions, such as including the 
entire absorption line, resulted in large and unstable changes in cross-correlation velocities with 
minor changes in fitting parameters, while smaller regions could exclude a signal from a high 
velocity amplitude binary.  

We cross-correlated each spectrum with three different template spectra: the coadded ESI spectrum 
for each WD, the highest single signal-to-noise object spectrum of each WD, and the Koester model 
DA spectrum corresponding to the atmospheric parameters from \citet{Williams2007}.  We resampled these model 
spectra to vacuum wavelengths and smoothed by a Gaussian kernel with FWHM equal to our 
spectral resolution.   This final template option resulted in the smallest uncertainties in velocity as calculated by \emph{fxcor}; these errors of $\approx 7$ 
km s$^{-1}$, were verified by the standard deviation of the measurements: $7.0$ km s$^{-1}$ and 6.5 km s$^{-1}$ for LAWDS 4 and LAWDS 7, respectively, under the 
assumption of constant radial velocities.  This uncertainty is also similar to our expectations based 
on the instrumental resolution and object spectrum signal-to-noise.   We therefore chose to use the 
model spectra as our cross-correlation templates.  

The Koester model spectra predate the recent changes to physical parameters in model WD 
atmospheres, so we tested for potential effects of template mismatch by cross-correlating the 
observed spectra with models of significantly different \teff and \logg.  We found systematic offsets of a few km/s, 
comparable in size to our stated measurement errors.  Further, since the updated physics does not affect the rest 
wavelength of the line centers, and since these centers dominate the cross-correlation signal, we 
are confident that our conclusions would not change significantly with newer models.  

\begin{figure}
\begin{center}
\includegraphics[angle=270,width=0.99\columnwidth]{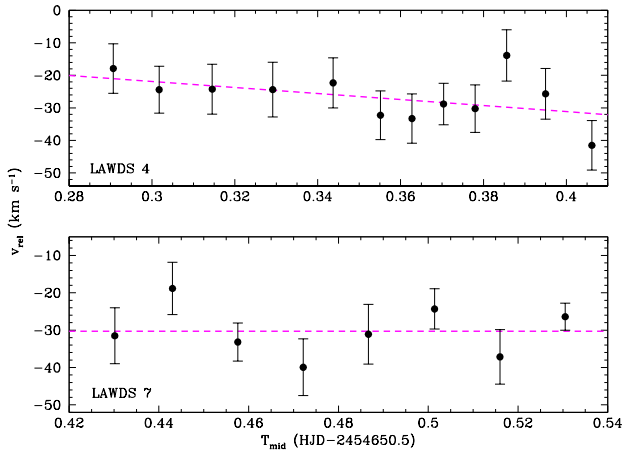}
\end{center}
\caption{Relative velocities for LAWDS 4 (top) and LAWDS 7 (bottom) after cross-correlation with the best-fitting model spectra.  The dashed magenta line indicates the best-fitting linear trend for LAWDS 4 and the mean velocity for LAWDS 7.  Neither object shows evidence of high-amplitude, short-period radial velocity variations.\label{fig.rvs}}
\end{figure}

The cross-correlation velocities are given in Table \ref{tab.wb4} for LAWDS 4 and Table \ref{tab.wb7} for 
LAWDS 7; these velocities are plotted in Figure \ref{fig.rvs}.  These are relative, not absolute, radial velocities.  Relative heliocentric corrections reflecting the change in velocity over the observations have been applied, but gravitational redshifts and pressure shifts of the H$\beta$ line centroid \citep[e.g.,][]{1987ApJ...313..750G,2015ApJ...808..131H} have been ignored at this point, since any orbital motion will be reflected as a change in radial velocity, and the gravitational redshift and pressure shifts should not change sign or magnitude due to orbital motion. 

 Neither LAWDS 
4 nor LAWDs 7 exhibits any evidence of high-amplitude short-period radial velocity variations.  The individual velocity measurements for LAWDS 7 are completely consistent with the mean relative velocity of $v_{\rm rel}=-30.3$ km s$^{-1}$ once the measurement errors are considered.  LAWDS 4 may show some evidence of a negative velocity trend over the observations, with a formal slope of $-93\pm 54$ km s$^{-1}$ d$^{-1}$, though this slope is not statistically significant.  If we assume that the observed slope is real and due to binary motion, the lack of any flattening or inflection of the radial velocity curve constrains any orbital period to be $> 5.5$ h, twice the length of the observational series.  

We determined the absolute radial velocities by measuring the centroid of the H$\alpha$ NLTE line 
core in the coadded spectrum of each object, as this measurement is unaffected by pressure shifts 
\citep{1987ApJ...322..852K,2010ApJ...712..585F,2015ApJ...808..131H}. Based on the spectral 
signal to noise ($\sim 100$ per resolution element for the summed spectra), we estimated the errors 
of the centroiding procedure to be $\approx 5.6$ km s$^{-1}$.  The velocities were then corrected to 
heliocentric velocities and listed in Table \ref{tab.mean}.

\begin{deluxetable}{lccccc}
\tablecolumns{4}
\tablecaption{Co-added spectrum velocities and gravitational redshift corrections\label{tab.mean}}
\tablewidth{0pt}
\tablehead{\colhead{Object} & \colhead{$v_{\rm obs}$} & \colhead{$v_{\rm GR}$}  & \colhead{$v_r$} \\ 
& (km s$^{-1}$)  & (km s$^{-1}$) & (km s$^{-1}$)  }
\startdata
LAWDS 4 & $36.1\pm 5.6$ & $48.3\pm 3.7$ & $-12.2\pm 6.7$ \\
LAWDS 7 & $22.4\pm 5.6$ & $58.3\pm 4.4$ & $-35.9\pm 7.1$ \\
\enddata
\tablecomments{Velocities are solely for the H$\alpha$ NLTE line core to avoid uncertainties due to pressure shifts in the H$\beta$ line.  Errors in the gravitational redshift are calculated from mass uncertainties in Table \ref{tab.newfits}.  The errors in $v_r$ are these $v_{\rm obs}$ and $v_{\rm GR}$ errors added in quadrature. }
\end{deluxetable}

\section{Discussion}
For each WD, we estimated the minimum expected radial velocity changes over the course of the observations by a DD Type Ia supernova progenitor system.  If we assume that the total system mass is $1\mch$ and the merger timescale is 14 Gyr, and if we assume that the observations are centered precisely on the maximum or minimum of the radial velocity curve, then we would expect a change in radial velocity of $\gtrsim 40 \sin i$ km s$^{-1}$.  Given the lack of significant velocity changes, LAWDS 4 and/or LAWDS 7 therefore can only be Type Ia supernova progenitor systems if we are observing a long-delay progenitor at low inclination and at the least favorable orbital phase.

Could LAWDS 4 and LAWDS 7 simply be wide binary systems in NGC 6633 with orbital timescales and velocity amplitudes not detectable in our one night of observations?  \citet{1997MNRAS.292..177J} and \citet{2013A&A...558A..53K} found the heliocentric radial velocity of NGC 6633 is $\approx -28$ km s$^{-1}$.  In order to compare our radial velocities for LAWDS 4 and LAWDS 7 to the cluster velocity, we must correct the measured velocities for gravitational redshift.

As pointed out by, e.g.,  \citet{2010ApJ...712..585F}, the measured velocity $v_{\rm obs}$ is the sum of a star's radial velocity  $v_r$ and the gravitational redshift $v_{\rm GR}$; i.e., $v_{\rm obs} = v_r + v_{\rm GR}$.  We calculate the predicted gravitational redshift from the revised \logg and mass of each WD (see Section \ref{sec.params}).  The calculated $v_{\rm GR}$ and derived radial velocity $v_r$ for both WDs are given in Table \ref{tab.mean}. 

The corrected WD heliocentric radial velocity of LAWDS 7 is consistent with the cluster radial velocity at the $1.1\sigma$ level, while that of LAWDS 4 differs at the $2.5\sigma$ level.  Therefore, LAWDS 4 is unlikely to be a cluster member. 

As for LAWDS 7, in \citet{Williams2007} we noted that its color ($\bv=0.23\pm 0.03$) was  too red based on the reddened best-fit spectral models ($\bv=0.15$ for our revised parameters in this paper).  This mismatch, if significant, could be due to the presence of a cooler companion\footnote{The text of \citet{Williams2007} incorrectly compared the model prediction to the photometry of \object{LAWDS NGC 6633 6}.}.  Single WD model spectra can give satisfactory fits to composite WD  spectra  \citep{1989ApJ...345L..91B,2007ApJ...667.1126L}, and both H$\alpha$ NLTE cores in a long-period or low-inclination binary should be near the same systemic velocity and thus indistinguishable at our spectral resolution, so we conservatively conclude that LAWDS 7 remains a viable candidate cluster member DD system.   

Ultimately, parallax data from GAIA should be of sufficient quality to settle the question of binarity for both WDs.  If one or both WDs are truly cluster members, over-luminous, and therefore likely binaries, this would support our initial conjecture that intermediate-age open clusters may indeed be fruitful hunting grounds for massive WD binaries.

\clearpage
\acknowledgements 
K.A.W.~ is grateful for the financial support of National Science
Foundation award AST-0602288.   This work was done in part through the REU Program in Physics and Astronomy at Texas A\&M University-Commerce funded by the National Science Foundation under grant PHY-1359409.  S.C.~was supported in part by a Summer Research Assistant award from Texas A\&M University-Commerce.   We thank S.~Casewell, D.~Koester, and S.~Geier for useful insight provided in discussions at the 2014 and 2016 European White Dwarf Workshops.  We also thank M.~Wood for insightful discussions and his mastery of Python scripts used during analysis of some unfortunately spurious results.  K.A.W.~ also wishes P.~Dobbie well, thanks him profusely for many important and insightful discussion, and will miss his careful and important contributions to the field.
Keck telescope time was granted by NOAO, through the Telescope System Instrumentation Program (TSIP). TSIP is funded by NSF.
The authors wish to recognize and acknowledge the very significant cultural role and reverence that the summit of Maunakea has always had within the indigenous Hawaiian community.  We are most fortunate to have the opportunity to conduct observations from this mountain. This research has made use of NASA's Astrophysics Data System and of the SIMBAD database, operated at CDS, Strasbourg, France.

{\it Facilities:} \facility{Keck:II (ESI,TSIP)}

%\bibliographystyle{../apj}
%\bibliography{../wd_jabref}

%\end{thebibliography}

\end{document}